\documentclass{aa}
\usepackage{graphicx}
\begin{document}

   \thesaurus{06     % A&A Section 6: Form. struct. and evolut. of stars
              (02.14.1;  % Nuclear reactions, nucleosynthesis,abundances
               08.23.2)}  % Stars: Wolf-Rayet,
   \title{Synthesis of $^{19}$F in Wolf-Rayet stars}

%   \subtitle{  }

   \author{G.~Meynet
          \inst{1}
          \and
          M.~Arnould\inst{2}\fnmsep
          }

   \offprints{G.~Meynet}

   \institute{Geneva Observatory, CH-1290 Sauverny, Switzerland\\
              email: georges.meynet@obs.unige.ch
         \and
             Institut d'Astronomie et d'Astrophysique, Universit\'e  Libre de
Bruxelles,
 Campus Plaine, CP 226, B-1050 Brussels, Belgium\\
             email: marnould@astro.ulb.ac.be
             }

   \date{Received ; accepted  }

   \maketitle
% DECLARATIONS FROM THE AUTHORS
%
\newcommand{\chem}[2]{$\rm{}^{#1}\kern-0.8pt#2$}
\newcommand{\chim}[2]{\rm{}^{#1}\kern-0.8pt#2}
\newcommand{\reac}[6]{$\rm\,{}^{#1}\kern-0.8pt{#2}\,({#3}\,,{#4})\,
           {}^{#5}\kern-0.8pt{#6}\,$}
\newcommand{\gsimeq}{\,\,\raise0.14em\hbox{$>$}\kern-0.76em\lower0.28em\hbox  
{$\sim$}\,\,}
\newcommand{\lsimeq}{\,\,\raise0.14em\hbox{$<$}\kern-0.76em\lower0.28em\hbox  
{$\sim$}\,\,}
\newcommand{\ms}{\rm M_{\odot}}
\newcommand{\zs}{\rm Z_{\odot}}
\newcounter{obj}
\newcounter{cnt}

\begin{abstract}

Meynet \& Arnould (1993a) have suggested that Wolf-Rayet (WR) stars could
significantly conta\-minate the Galaxy with \chem{19}{F}. In their scenario,
\chem{19}{F} is synthesized at the beginning of the He-burning phase from
the \chem{14}{N} left over by the previous CNO-burning core, and is ejected
in the interstellar medium when the star enters its WC phase. Recourse to CNO
seeds
makes the \chem{19}{F} yields metallicity-dependent.  

These yields are calculated on grounds of detailed stellar evolutionary
sequences 
for an extended range of initial masses (from 25 to 120 $\ms$) and
metallicities ($Z=0.008$, 0.02 and 0.04). The adopted mass loss rate
prescription
enables to account for the observed variations of WR populations in different
environments. 

The \chem{19}{F} abundance in the WR winds of 60 M$_\odot$ model 
stars is found to be about 10 to 70 times
higher than its initial value, depending on the metallicity. 
This prediction is used in conjunction with a
very simple model for the chemical evolution of the Galaxy to predict that WR
stars could be significant (dominant ?) contributors to the solar system 
fluorine content. We also briefly discuss the implications of our model
on the possible detection of fluorine at high redshift.

\end{abstract}

%
%________________________________________________________________

\section{Introduction}

For long, the solar system has been the only location of the Galaxy with a
known fluorine (\chem{19}{F}) abundance. At the same time, the production 
site(s) of
this element has been a major nucleosynthetic puzzle, even if F is the least 
abundant (mass fraction of $4\,10^{-7}$, following Grevesse \& Sauval 1998) 
of the elements ranging from carbon to calcium. 

These last years, the 
situation has changed quite dramatically, both observationally and 
theoretically. Fluorine overabundances (with respect to solar) in MS, S and C 
stars have been reported (Jorissen et al. 1992), and correlate in 
particular with s-process enrichments. These observations demonstrate that
thermally pulsating Asymptotic Giant Branch (AGB) stars are fluorine 
producers, as
predicted by Goriely et al. (1989), and confirmed by calculations conducted
in the framework of detailed AGB models (Forestini et al 1992, Mowlavi et al.
1996, 1998). It remains of course to determine the exact level of the
contribution of 
these (mass losing) stars to the solar system and galactic F content.

In direct relation with this question, various calculations have been made in
order to estimate the \chem{19}{F} yields from massive stars. The neutrino 
process operating during supernova
explosions has been envisioned as a possible producer of primary \chem{19}{F}
(e.g.
Woosley \& Weaver 1995). On the other hand, Meynet \&
Arnould (1993a) have investigated on grounds of detailed stellar models the
suggestion (Goriely et al. 1989) that the
hydrostatically burning He-shell can synthesize \chem{19}{F} of secondary
nature. They
find that the
level of production is relatively modest in $M \lsimeq 20\,\ms$. In contrast,
they show that stars which are massive enough to become Wolf-Rayet (WR) stars
can
eject through their winds substantial amounts of fluorine synthesized in the
core at the beginning of the He-burning phase. 

In the present work, we revisit the
question of the galactic contribution of WR stars to 
\chem{19}{F} with the help of new stellar models that better account for many
important observable properties of WR stars. In addition, we extend the range 
of masses and metallicities considered in our previous study. The broadening of
the
explored metallicity range may take some additional importance in relation with
the
recent claim by Timmes et
al. (1997) that ``{\it positive detection of any fluorine at a sufficiently
large redshift
($z \gsimeq 1.5$) would suggest strongly a positive detection of the neutrino
process operating in massive stars}''. The possibility of a significant
thermonuclear production of \chem{19}{F} by WR stars of different metallicities
might
blur this picture, and might at least imply the necessity of esta\-blishing
observationally the primary or secondary nature of the detected fluorine, if
any.

The physical ingredients of the models are discussed in Sect.~2. 
Section~3 presents our predicted yields from individual WR stars, while Sect.~4
gives a rough estimate of the contribution of WR stars to the galactic
\chem{19}{F} content. Some conclusions are drawn in Sect.~5.
 
\section{The physical ingredients of the stellar models}

The evolutionary models are computed with the same physical ingredients
as in Meynet et al. (1994). However, the adopted nuclear reaction network is
extended,
especially in order to include the reactions involved in the production and
destruction of \chem{19}{F} (see below).
 
The differences with respect to the computations of Meynet \& Arnould (1993a)
are
twofold:
 
\noindent 1) A more extended range of initial masses (from 25 to 120 $\ms$)
and metallicities ($Z=0.008$, 0.02 and 0.04) is explored; 
 
\noindent 2) The present grid of models is computed with the mass loss rates
adopted by Meynet et al. (1994), which are twice as large as the values of  
$\dot M$ recommended by de Jager et al.
(1988) and Conti (1988) for the pre-WR and WNL phases. This mass loss rate
prescription
enables to account for the observed variations of WR populations in different
environments (Maeder \& Meynet 1994). 

The metallicity
dependence of the mass loss rates during the pre-WR phases is adopted from
previous
works (e.g. Meynet et al. 1994). More specifically, $\dot M$ scales
with metallicity $Z$ according to $\dot M_Z/\dot M_\odot=(Z/Z_\odot)^{0.5}$, 
where $Z_\odot$ is the solar metallicity. This scaling is deduced from
stellar wind models (cf. Kudritzki et al. 1987, 1991).

Let us finally add that the models are computed with a moderate core
overshooting
($d/H_p= 0.20$, where $d$ is the overshooting distance and $H_p$ the pressure
scale height at the boundary of the classical core). 

\subsection{The thermonuclear \chem{19}{F} production and destruction
paths}

The CNO mode of H-burning is responsible for the production and destruction 
of \chem{19}{F} through the reaction chain

$$^{14}{\rm N}(p,\gamma)^{15}{\rm O}(\beta^+)^{15}{\rm N}(p,\gamma)^{16}{\rm O}
(p,\gamma)^{17}{\rm F},$$
$$^{17}{\rm F}(\beta^+)^{17}{\rm O}(p,\gamma)^{18}{\rm F}(\beta^+)
^{18}{\rm O}(p,\gamma)
^{19}{\rm F}(p,\alpha)^{16}{\rm O}.$$

The adopted $^{19}{\rm F}(p,\alpha)^{16}{\rm O}$ rate is the geometrical mean 
of the lower and upper limits to that rate proposed by Kious (1990).

Fluorine can also be produced and destroyed during He-burning through the 
chains (see also Meynet \& Arnould 1993a) 
\vskip 15mm
\begin{eqnarray*}
                            &             &
(\beta^+)^{18}O(p,\alpha)^{15}N(\alpha,\gamma)^{19}F                \cr
                            & \nearrow    &  \hskip 4.2cm  \searrow            
                                  \cr 
^{14}N(\alpha,\gamma)^{18}F & \rightarrow &
(n,p)^{18}O(p,\alpha)^{15}N(\alpha,\gamma)^{19}F(\alpha,p)^{22}Ne.  \cr
                            & \searrow    &   \hskip 4.2cm  \nearrow           
                                  \cr
                            &             & \hskip 0.5cm
(n,\alpha)^{15}N(\alpha,\gamma)^{19}F                               \cr
\end{eqnarray*}

\noindent The synthesis of \chem{19}{F} thus requires the availability 
of neutrons and protons. They are mainly produced by the reactions
\reac{13}{C}{\alpha}{n}{16}{O} and \reac{14}{N}{n}{p}{14}{C}.

 The first chain of
transformation of \chem{14}{N} into \chem{19}{F} mentioned above is by far the
most
important in the conditions of relevance in this work, where the
$\beta^+$-decay lifetime $\tau_\beta$(\chem{18}{F}) of \chem{18}{F} is much
shorter
than its lifetime $\tau_{n,\alpha}$(\chem{18}{F})  or
$\tau_{n,p}$(\chem{18}{F})
against  ($n,\alpha$) or ($n,p$) reactions. For example, $\tau_\beta$ is a few
hours
only at the center of a 60 M$_\odot$ model at the beginning of core He-burning,
while
the corresponding  $\tau_{n,\alpha}$(\chem{18}{F}) and
$\tau_{n,p}$(\chem{18}{F})
amount to about 1~400 and 18~000 years, respectively. 

The NACRE compilation of reaction rates (Angulo et al. 1999) was not available
yet
at the time of completion of the calculations reported here. This is why most
of the necessary nuclear data are taken from Caughlan \& Fowler (1988). There
are
some exceptions to this rule. In particular, the \chem{13}{C} $\alpha$-capture
rate is taken from Descouvemont (1987),  whose theoretical prediction of an
increase of the astrophysical S-factor at low energies is confirmed
experimentally (see NACRE). 
The \reac{14}{N}{n}{p}{14}{C} rate is taken from Brehm et al. (1988). It is a 
factor of two lower than the one proposed by Koehler and O'Brien (1989), and
leads
consequently to a {\it lower limit} of the calculated \chem{19}{F} yields.

\section{Predicted \chem{\bf 19}{\bf F} yields from individual WR stars}

\begin{figure*}
\begin{center}
\resizebox{15cm}{!}{\includegraphics{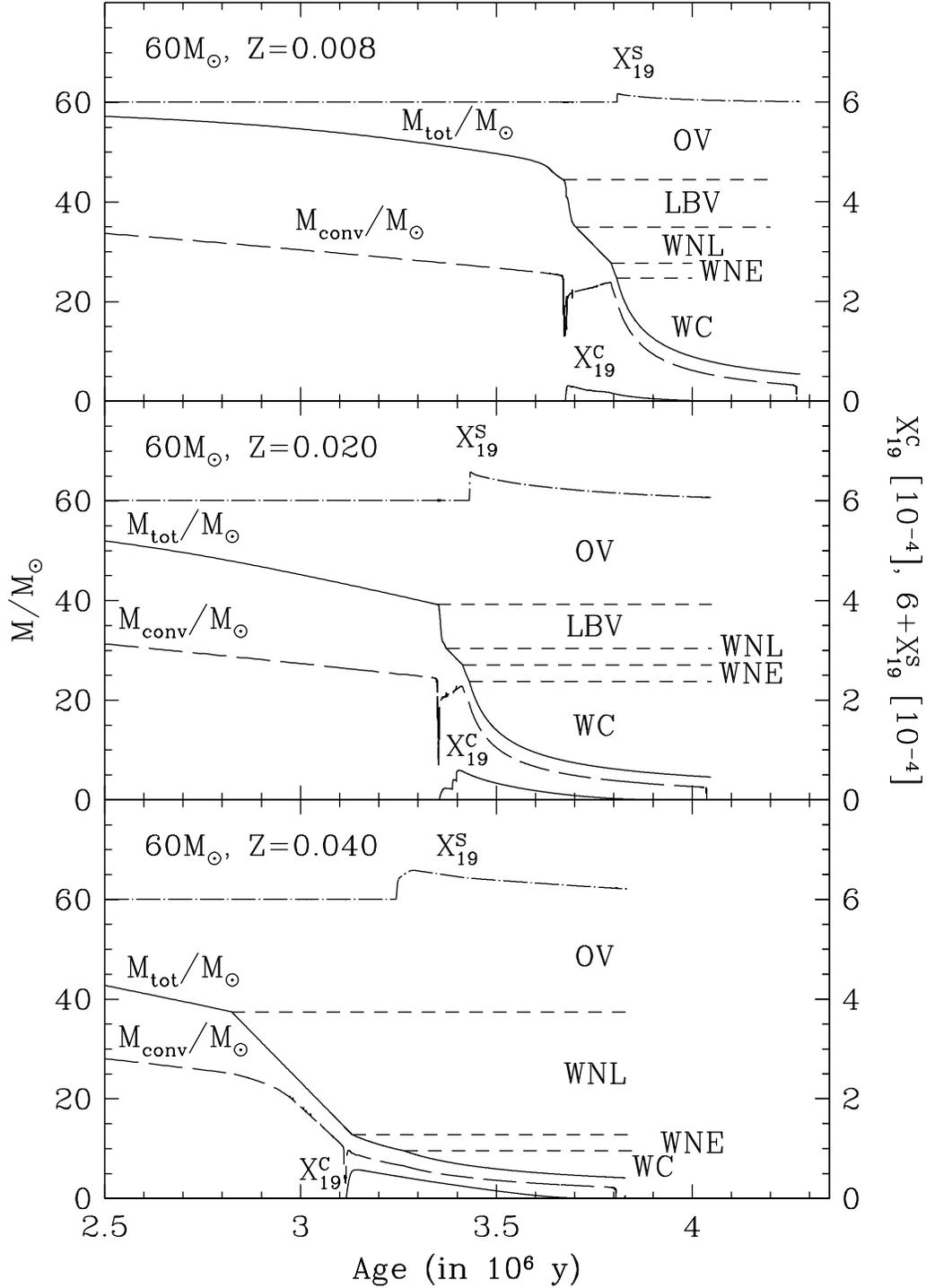}}
\end{center}
\caption{Evolution of the total mass M$_{\rm tot}$, of the mass of the 
convective
core M$_{\rm conv}$, and of the central ($X_{19}^{\rm c}$) and surface 
($X_{19}^{\rm s}$) \chem{19}{F} mass fractions for the 60 M$_\odot$ model stars
with metallicities $Z = 0.008$, 0.020 and 0.040 during the end of the H-burning
stage and the whole He-burning phase. The initial \chem{19}{F} mass
fraction is assumed to relate  to the solar value $X_{19}^{\odot}$ by
$X_{19}^0(Z)=(Z/Z_\odot)
X_{19}^{\odot}$. The spectroscopic types encountered during the evolution 
are indicated on the right of the figure: OV for O-type main sequence stars,
LBV for Luminous Blue Variables, WNL, WNE and WC for the different
classes of WR stars. Note the 
different ordinate scales on the left and on the right of the figure.}
\label{Figun}%  
\end{figure*}

As discussed by Arnould et al. (1999) on grounds of the NACRE rates,
\chem{19}{F}
could be overproduced (with respect to solar) by the CNO cycle only at
temperatures 
around $15 \times 10^6$ K, the exact level of this overproduction remaining
poorly
predictable, however, in view of remaining rate uncertainties. This conclusion
contradicts the one derived from the use of the rates recommended by Caughlan
\&
Fowler (1988), in which case fluorine can never emerge in significant  amounts
from the CNO burning. As the latter rates are adopted in our calculations, the
CNO
zones of the computed model stars are depleted in \chem{19}{F}. This translates
directly into a decrease of  the \chem{19}{F}  mass fraction $X_{19}^{\rm s}$
at
the stellar surfaces  when the 
\chem{19}{F}-depleted CNO ashes are uncovered by mass loss (with the choice
of the ordinate scales, the changes of fluorine abundance at the center
and at the surface during the H-burning phase are not visible on Fig.~1). With
the NACRE rates, it is expected that more \chem{19}{F} would be present at the
surface. However, it is also likely that this change is not able to affect
drastically the predicted final yields, as these are dominated by the
\chem{19}{F}
made during the He-burning phase.

In fact, as seen in Fig.~1, fluorine
builds up through  $^{14}{\rm N}(\alpha,\gamma)^{18}{\rm F}(\beta^+)^{18}{\rm
O}
(p,\alpha)^{15}{\rm N}(\alpha,\gamma)^{19}{\rm F}$ during the early phase of
core
He-burning. However, at the end of He-burning, \reac{19}{F}{\alpha}{p}{22}{Ne}
is
responsible for a significant 
\chem{19}{F} destruction.
Thus, material experiencing the whole He-burning episode cannot be
\chem{19}{F}-enriched. 
{\it In contrast, in massive stars going through the WR stage} 
(initial mass $M_{\rm i} \gsimeq 25\,\ms$ for $Z=0.02$, $M_{\rm i} \gsimeq
35\,\ms$ for $Z=0.008$;
see Maeder \& Meynet 1994), {\it some \chem{\rm 19}{\rm F} synthesized early
during the
core He-burning  phase is ejected into the interstellar medium by stellar winds
before its destruction}. Indeed, Fig. 1 exhibits an increase of $X_{19}^{\rm
s}$  when the He-burning products  appear at the surface during the WC phase.
As a result, the ratio
$\langle X_{19}^{\rm s}(\rm {WC})\rangle/X_{19}^{\odot}$ of the average 
\chem{19}{F} surface mass fraction during the whole WC phase to the solar 
system \chem{19}{F} mass fraction takes values as high as about 55, 95 and 60 
in the case of the $60\,\ms$ model stars with $Z = 0.008$, 0.02 and 0.04, 
respectively.

Figure 2 shows the \chem{19}{F} ``wind'' yields for the computed stars
($M_{\rm i}$, $Z$) with initial mass $M_{\rm i}$ and metallicity $Z$. These 
yields, noted $p^{\rm wind}_{19}(M_{\rm i},Z)$, are equal to
 
$$p^{\rm wind}_{19}(M_{\rm i},Z)=$$
$$\int_0^{\tau(M_{\rm i},Z)} \dot M(M_{\rm i},Z,t)
 [X_{19}^{\rm s}(M_{\rm i},Z,t)-X_{19}^0(Z)] dt,\eqno (1)$$

\noindent where $\tau(M_{\rm i},Z)$ is the total lifetime of the 
star ($M_{\rm i}$, $Z$), $\dot M(M_{\rm i},Z,t)$ its mass loss rate at 
age $t$, $X_{19}^{\rm s}(M_{\rm i},Z,t)$ its \chem{19}{F} surface mass 
fraction at age $t$,
and $X_{19}^0(Z)$ its initial \chem{19}{F} mass fraction, assumed to relate 
to $X_{19}^{\odot}$ by $X_{19}^0(Z)=(Z/Z_\odot) X_{19}^{\odot}$.
These yields may be negative in case most of the ejected material has
been depleted in fluorine.

Figure 2 demonstrates that the highest yields are obtained for stars
with $Z = Z_{\odot}$ and $40 \lsimeq M_{\rm i} \lsimeq 85\,\ms$. At lower 
metallicities, the winds are indeed weaker, and thus uncover the
He-burning core only for the most massive stars and when
the $^{19}$F has already been burnt. On the other hand,
at higher metallicities and for $M_{\rm i} \gsimeq 85\,\ms$, the H-burning
core mass decreases so rapidly during the main sequence as a consequence of
very strong stellar winds that the He-burning
core becomes too small for being uncovered by the stellar winds.

The above discussion shows that the most important physical ingredient
influencing the WR \chem{19}{F} yields is the metallicity-dependent mass loss
rates, quantities like convective core masses being less crucial in this
respect. As
a numerical example, the value for $\langle X_{19}^{\rm s}(\rm
{WC})\rangle/X_{19}^{\odot}$ rises from about 18 in the 60 M$_\odot$ low mass
loss rate
model of Meynet \& Arnould (1993a) to about 95 in the same model star computed
in this
paper with an increased $\dot M$ value. This high sensitivity to $\dot M$ might
cast doubts on the reliability of the predicted \chem{19}{F} yields. In fact,
some
confidence in the results presented in this paper may be gained by noting that
our
present choice of the mass loss rates allows to account for the variation with
metallicity of the number ratio of WR to O-type stars in regions of constant
star
formation rate (Maeder \& Meynet 1994).
 
\begin{figure}
\begin{center}
\resizebox{9cm}{!}{\includegraphics{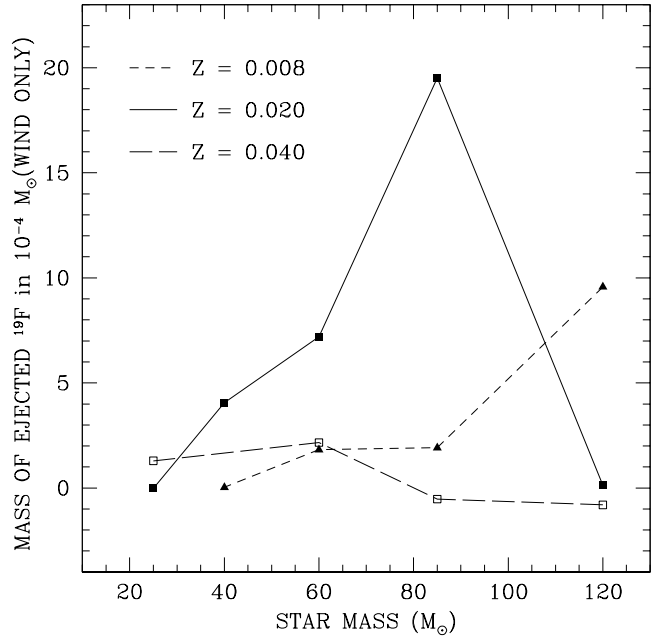}}
\end{center}
\caption{Mass of \chem{19}{F} ejected by the stellar winds ($p_{19}^{\rm wind}$
in Eq. 1) as a function of the initial mass and metallicity.}
    \label{Figci}%
    \end{figure}

\section{Estimate of the contribution of WR stars to the galactic fluorine} 
 
In order to evaluate the level of \chem{19}{F} contamination by the winds of WR
stars on a galactic scale, we use the $p_{19}^{\rm wind}$ yields [Eq. (1)] in 
a very 
simple model of galactic chemical evolution making use of the closed box and
instantaneous recycling approximations. We also suppose that
only WR stars are able to affect the galactic \chem{19}{F} budget through 
their winds, all other possible production or destruction sites being 
neglected. 

In such conditions, the \chem{19}{F} mass fraction $X_{19}(t)$ in the galactic
gas at time $t$ is equal to (e.g. Tinsley 1980)
$$X_{19}(t)= \widetilde{y_{19}} \ln [1/\sigma (t)],\eqno (2)$$
where {\bf $\sigma (t)$} is the mass fraction of the gas in the Galaxy 
at time $t$, and $\widetilde{y_{19}}$ is a representative time-independent
approximation of the net yield of a stellar generation defined by
$$y_{19}(t) ={1 \over 1-R} \int_{M_1}^{M_2}
p^{\rm wind}_{19}(M_{\rm i},Z(t)) \Phi(M_{\rm i}) dM_{\rm i},\eqno (3)$$
where $R$ is the ``returned fraction'', $M_1$ and $M_2$ the lowest and
highest mass of the stars going through the WR phase, and 
$\Phi(M_{\rm i})$ the initial mass function (IMF). It has to be noted 
that Eq. (2) would break down if the true time- (or $Z$-)dependent $y_{19}(t)$
yields were used instead of $\widetilde{y_{19}}$. In order to evaluate the
latter quantity, we notice that $y_{19}(Z(t))$ values of about
$10^{-7}$, $5.7\ 10^{-7}$ and $2.2\ 10^{-7}$ are obtained for 
$Z = 0.008$, 0.02 and 0.04 if use is made of the $p_{19}^{\rm wind}$ values 
reported in Sect.~3 and of
the (properly normalized) IMF derived by Kroupa et al. (1993). On such 
grounds, 
we just adopt the rough estimate $\widetilde{y_{19}} \approx 3\,10^{-7}$.

If this approximation is used in conjunction with the value 
$\sigma \approx 0.2$ considered to characterize the solar 
neighbourhood at the time of the solar system formation 4.5 billion years ago 
(see Prantzos \& Aubert 1995, and references therein), 
Eq.~(2) leads to 
$X_{19} \approx 5\,10^{-7}$ in the local $Z = Z_{\odot}$ interstellar medium
(to be compared with the solar system abundance of 4~10$^{-7}$). 
Thus, our simple estimate predicts that {\it WR stars might account for most of
the solar system} \chem{19}{F} {\it content}.  Even larger
\chem{19}{F} quantities would be predicted with the use of the 
\reac{14}{N}{n}{p}{14}{C} rate of Koehler \& O'Brien (1989)! After having faced
for
long the problem of the underproduction of \chem{19}{F}, the theory of
nucleosynthesis
might now live with the danger of its predicted overabundance. If this is
confirmed by
further studies, constraints will obviously have to be put on one model or
another.
 
\section{Implications of \chem{\bf 19}{\bf F} detection at high redshift}

Any \chem{19}{F} present at high redshifts has to have been synthesized in
massive stars only. Timmes et al. (1997) have argued further that its detection
at
redshifts $z \gsimeq 1.5$ would in fact be a signature of the $\nu$-process in
massive
star explosions. The possibility of \chem{19}{F} production by non-exploding WR
stars
might in fact weaken this statement, and blur the picture substantially. 

Of course, one has to acknowledge that the contribution from WR stars at high
redshifts
may be reduced as a direct result of the lower metallicities that appear to
cha\-racterize such regions. According to observations of
Damped Lyman $\alpha$ systems (Pettini et al. 1997), the metallicity at
redshifts between 1.5 and 2
indeed lies around $0.1 Z_\odot$.  Such a reduced metallicity lowers the WR
\chem{19}{F} yields for two reasons. First, the number of WR stars predicted by
non-rotating single star models is considerably reduced as a result of lower
mass
losses (Maeder \& Meynet 1994). Second, the abundances of the CNO seeds that
are needed for the secondary WR \chem{19}{F} production are reduced as well.

Even so, it would certainly be premature at this point to completely forget
about the
role of WR stars in a possible enrichment of high-$z$ material with
\chem{19}{F}, and
to relate it strictly with the $\nu$-process.This is even more true as the
predictions
reported in this paper are based on single, non-rotating stellar
models only. How binarity and/or rotation would change these results remains
to be checked. At present, the published rotating evolutionary models
leading to WR stars  (Fliegner \& Langer 1994, Meynet 1998, 1999) make
no predictions concerning the synthesis of fluorine. However they show that
rotation favours an early entrance into the WR phase for a given mass,
and decreases the minimum initial mass for a star to go through a WR phase at a
given metallicity. Moreover, the mixing induced by rotation
opens up new nucleosynthetic channels (see Heger 1998) whose
importance for the scenario of fluorine production presented in this
paper remains to be quantitatively assessed. Finally, let us note that the
effects of rotation might be more important at low $Z$
if, as suggested by Maeder et al. (1999), the average rotation is faster at low
metallicities.
In such conditions, and in
absence of quantitative calculations, one has to remain alert to the
possibility  of a
significant contamination of low metallicity high redshift regions by the
\chem{19}{F}-loaded wind of WR stars.   

Clearly, observations of \chem{19}{F} at high redshift, if  possible at all,
would be
decisive in order to answer the question of the very production mechanism of
this
element. An important distinguishing feature would be the primary nature of the
observed \chem{19}{F}, as predicted by the $\nu$-process, or its secondary
behaviour,
as expected from the thermonuclear model discussed in this paper.

\section{Conclusion}

Detailed stellar model predictions made in the framework of a very
rough model for the chemical evolution of the solar neighbourhood leads to the
conclusion  
that non-exploding non-rotating single WR stars alone could account for the
solar
\chem{19}{F} content. This conclusion remains to be ascertained by the adoption
of a more realistic galactic evolution model. Still, it appears likely
that the considered WR stars might be significant, and even possibly dominant,
galactic
\chem{19}{F} contributors. In addition, they might well be responsible for a
\chem{19}{F} enrichment, if any, of high-redshift ($z \gsimeq 1.5$)
low-metallicity regions ($\sim$ 0.1 $Z_\odot$).
Further
predictions are eagerly awaited for rotating, as well as binary, WR stars.
 
Finally, let us stress that the most direct test of the present model would
be the measurement of the abundance of fluorine in the wind of WC stars. 
It remains to be seen if such observations are really feasible. 

\begin{acknowledgements}
We thank N. Prantzos for comments on the galactic chemistry aspects of this
work. This research
has been supported in part by the HCM Programme of the European Union
under contract ERBCHRXCT 930339. 
\end{acknowledgements}

\end{document}